\documentclass[runningheads]{llncs}
\usepackage{graphicx}
\usepackage{amsmath,amssymb,amsfonts,comment}
\usepackage{multirow}
\usepackage{lipsum,booktabs}
\usepackage{caption}
\usepackage{subcaption}
\usepackage{hyperref}
\usepackage{adjustbox}
\usepackage{color, soul}
\sethlcolor{yellow}

\begin{document}
\title{Knowledge-Distilled Graph Neural Networks for Personalized Epileptic Seizure Detection \thanks{This work is supported by the PEDESITE project\\${}^2$This work was done by the third author when she was a research intern in the same lab as the other authors at EPFL. }}
\titlerunning{ }
%
 \author{Qinyue Zheng\inst{1} \and
 Arun Venkitaraman\inst{1} \and
 Simona Petravic\inst{2} \and 
 Pascal Frossard\inst{1}}
 \authorrunning{Q. Zheng, A. Venkitaraman, S. Petravic, and P. Frossard}
\institute{Signal Processing Laboratory LTS4, EPFL, Lausanne, Switzerland\and
 Embark Studios, Stockholm, Sweden \\
 \email{qinyue.zheng@epfl.ch,
 arun.venkitaraman@gmail.com, petravic.s@gmail.com,
 pascal.frossard@epfl.ch }\\}
%
\maketitle              
\begin{abstract}
Wearable devices for seizure monitoring detection could significantly improve the quality of life of epileptic patients. However, existing solutions that mostly rely on full electrode set of electroencephalogram (EEG) measurements could be inconvenient for every day use. 
In this paper, we propose a novel knowledge distillation approach to transfer the knowledge from a sophisticated seizure detector (called the teacher) trained on data from the full set of electrodes to learn new detectors (called the student). They are both providing lightweight implementations and significantly reducing the number of electrodes needed for recording the EEG. We consider the case where the teacher and the student seizure detectors are graph neural networks (GNN), since these architectures actively use the connectivity information. We consider two cases (a) when a single student is learnt for all the patients using pre-selected channels; and (b) when personalized students are learnt for every individual patient, with personalized channel selection using a Gumbel-softmax approach.
Our experiments  on the publicly available
Temple University Hospital EEG Seizure Data Corpus (TUSZ) show that both knowledge-distillation and personalization play significant roles in improving performance of seizure detection, particularly for patients with scarce EEG data. We observe that using as few as two channels, we are able to obtain competitive seizure detection performance. This, in turn, shows the potential of our approach in more realistic scenario of wearable devices for personalized monitoring of seizures, even with few recordings.

\end{abstract}
\section{Introduction}
\label{sec:intro}
Epilepsy is a neurological disorder that is characterized by recurring, unprovoked seizures caused by surges of electrical activity in the brain and affects nearly three million people \cite{epilepsy}. About one third of the patients do not respond to treatment by drugs \cite{stats}. Hence, real-time seizure monitoring is crucial for improving the patients' quality of life, for example, by alerting caregivers that their assistance is needed once a seizure occurs. A continuous monitoring of the electroencephalogram (EEG) is useful in identifying and even predicting seizures in critically ill patients \cite{eeg}, particularly with the use of deep-learning approaches \cite{siddiqui2020review,raghu2020eeg,ievsmantas2020convolutional,ahmedt2020neural,roy2020seizure} The monitoring is usually performed in a hospital environment over the course of several days, which makes it infeasible to monitor patients long-term {in non-ambulatory settings}. Wearable devices could overcome the need of specialised intrusive medical equipment and hospital environment and enable real-time seizure monitoring on a daily basis. Existing measurement devices \cite{cap} that use EEG head caps with over 20 wired electrodes are however uncomfortable and difficult to wear over prolonged intervals and lighter and more discrete wearables are desirable for patients. 
Previous studies have attempted to reduce the number of EEG electrodes needed for seizure detection \cite{furbass_automatic_2017,eglass,DBLP:journals/artmed/GabeffTZCRRA21} with promising results. 
However, these solutions typically involve training detection systems from scratch for the new setting, and fail to incorporate the already existing historical EEG data of the patient recorded with many electrodes. Due to the nature of the disorder itself, seizure data is sparse in the number of available seizures and difficult to collect{, and it is thus important }to meaningfully use previous data. Further, it is known that the signals from the different regions of the brain (captured through the EEG electrodes) are not independent and exhibit strong inter-channel dependencies that could be viewed as a brain graph or a network. Hence, we ask the question:

\begin{center}
    {\it How to transfer information gained from a full set of channels/graph to {settings with a }reduced number of channels/subgraph while actively using the connectivity information? }
\end{center}

In this paper, we address this question by developing a novel approach for knowledge distillation (KD) with graph neural networks (GNNs) applied to seizure detection. Our motivation for the use of GNNs comes from the observation that they have been used extensively in applications with graph-structured data, and more recently have shown to result in promising seizure detection performance \cite{rahmani2023metagnn,selfsupervised}. More specifically, we propose a seizure detection model that consists of three interconnected blocks. Firstly, we have the knowledge distillation block, whereby we transfer the knowledge from a pre-trained seizure detection model to obtain a model that is light-weight and uses only a reduced set of input channels and the corresponding subgraph. Secondly, a channel selection block, which takes the full multi-channel input and retains the signal only on a reduced set of channels that are either pre-selected or learnt in a fully data-driven manner. Lastly, we have the GNN based seizure detection model that classifies the input in the form of the multi-channel signal from a reduced set of channels/electrodes and the corresponding subgraph, into seizure or non-seizure segments. 

Our goal is to also investigate the influence of two important aspects in seizure detection performance with reduced channels: (i) prior knowledge (through the use of the teacher model), and (ii) personalization/ patient-specific detection. 
The specific contributions of our paper are as follows:
\begin{itemize}
    \item We propose new GNN models for epileptic seizure detection that build on knowledge distillation to generate models that are both light-weight and work on subgraphs of reduced nodes/channels. To the best of our knowledge, this is the first KD approach dedicated to obtaining subgraph GNNs with reduced channels.
    \item We propose two different models for seizure detection with reduced channels, namely one with pre-selected (clinically motivated) channels and one with data-driven channels obtained from Gumbel softmax channel selection.
    \item By applying our approach on pre-trained GNN that uses a full electrode set, we obtain personalized (patient-specific) and global (non patient-specific) GNN models that are both lightweight (using only $\approx 3\%$ of the parameters of the teacher) and requires only a reduced subset of electrodes (requiring as low as only $10\%$ of the original electrodes)
    \item  We demonstrate the results of our approach on the TUH Seizure Corpus, which is one of the most popular and diverse datasets for epileptic seizures.
    \item We show empirically that the combination of personalization and KD could significantly improve seizure detection in cases of very scarce data, and in cases when the measurements are made from the relatively 'non-informative' electrodes. 
    
\end{itemize}
Finally, it could be noted that epilepsy seizure detection is a very active research problem. In particular, there has been a steady increase in the number of graph-based approaches, and particularly GNNs applied to the problem of seizure detection and classification \cite{selfsupervised,rahmani2023metagnn,covert2019temporal}. However, to the best of our knowledge no prior works exist that tackle the problem of channel reduction with GNNs and KD, particularly for seizure detection. While KD has been used in multiple settings related to GNNs \cite{graphkd1,graphkd2,graphkd3,graphkd4,graphkd5,graphkd6,graphkd7}, it has not been employed to the task of data-driven subgraph identification, which is the main objective in this paper. 

\section{Preliminaries}
\label{sec:preliminaries}
We now briefly review some of the basic concepts from GNNs and KD.

\textbf{Graph Neural Networks} 
Graph Neural Networks (GNNs) refer to a class of deep learning models designed for graph-structured data \cite{graphnn}. GNNs learn the representations of the nodes/channels in a graph and predict the labels or properties of nodes/edges by actively using the the underlying graph structure. Due to the graph structure, GNNs naturally provide an aspect of interpretability or explainability. GNNs have been shown to significantly outperform the use of CNNs or other non-graph approaches in many applications. While study and development of GNNs is an active research area, we consider the specific case of 
Graph convolutional networks (GCNs) in our work, since they form one of the simplest and most popular GNNs that directly generalize the convolution operation from CNNs to a graph setting \cite{gcn}. A multi-layer GCN has the layer-wise propagation rule in the hidden layers:
\begin{equation}
    \centering
H^{(l+1)}=\sigma({A}H^{(l)}\Theta^{(l)})
\end{equation}
where $H^{l}\in \mathbb{R}^{N \times D}$ is the hidden node features at $l$-th layer; $H^0$ denoting the input, $\sigma$ a non-linear activation function such as ReLU or sigmoid, $A$ the adjacency matrix, and $\Theta{(l)}$ being the weight matrix in the $l$-th layer that is learnt from the data for a given task.
Put simply, the graph convolution operation takes the weighted sum of the features of the neighbors of a node and applies a non-linear activation function to produce the updated features for the node. This operation is repeated for each layer, allowing the model to learn more complex representations of the graph structure and node features. The final output of a GCN is typically obtained by applying a linear layer to the features of the nodes in the final layer. Finally, depending on whether the task is regression or classification, the parameters of the GNN are learned by minimizing a loss function, respectively.

\textbf{Knowledge Distillation} 
Knowledge distillation (KD) \cite{hinton2015distilling} refers to transferring knowledge from a large/sophisticated pre-trained neural network (known as the {\em teacher} network) to a smaller network (known as the {\em student} network). The student represents a light-weight model derived from the teacher while enforcing the performance to be similar to that of the teacher.  A distillation loss is used during training to guide the student to replicate the teacher's behavior as closely as possible.
Different types of knowledge can be transferred, but the most straightforward one is response-based KD, which refers to the response of the output layer of the teacher. A widely used example of this is the class probability called as {\em soft targets} defined using a softmax function as
\begin{equation}
p(z_i, T) ={\exp(z_i/T)}/{\sum_{j}{}\exp(z_j/T)},    
\label{logits}
\vspace{-.1in}
\end{equation}
where $p_i$ is the probability of belonging to class $i$, $z$ is the vector of logits (outputs of the last layer of the teacher to a given input). The temperature $T$ controls the contribution of each soft target to the knowledge. When $T$ is equal to 1, we get the standard softmax function, but as $T$ increases, the probability distribution is softened. 
The distillation loss can be seen as comparing the class probabilities obtained from the teacher and the student.  It enforces the distribution of the outputs produced by the student to be close to that of the teacher. The Kullback-Leibler (KL) divergence is therefore often used as the distillation loss function, and minimizing this loss during training makes the logits of the student get closer to the logits of the teacher \cite{survey}. 
Let $z_t$ and $z_s$ denote the representation produced by the teacher and student models, respectively, for the same input.
Then, the final loss function used to train the student is a weighted average of the two terms and is defined as
\setlength{\arraycolsep}{0.0em}
\begin{align}
L_S{}={}&(1-\delta)L_D(p(z_t, T), p(z_s, T))
{+}\:\delta L_{CE}(y, p(z_s, 1)), 
\label{eq}
\end{align}
where $L_D$ is the distillation loss function, $p(z_t, T)$ are the teacher soft targets, $p(z_s, T)$ are the student soft targets, $L_{CE}$ is the cross entropy loss function, $y$ are the ground truth labels, and $\alpha$ is the weighting factor. The parameter $\delta$ represents the relative weight given to the teacher's knowledge over the new training data corresponding to the student training $-$ the higher $\delta$, the lesser the model relies on the teacher for the training of the student. We shall consider the KD as part of our approach later in Section \ref{sec:global_seizure_detection}.

\section{KD with GNNs for Seizure Detection}
\label{sec:global_seizure_detection}
\subsection{Proposed Model} We first propose our approach to design a global seizure detection student GNN that works on data with reduced nodes/channels and the corresponding subgraph, obtained using KD from a teacher GNN that operates on the complete node set. Let $D$ denote the number of nodes/channels in the full measurement. Let $A$ denote the adjacency matrix of the graph describing the connections between the different channels. The adjacency matrix could be obtained in different ways like a correlation matrix, functional connectivity, or simply the matrix that captures the physical proximity of the electrodes on the scalp. In our paper, we use the latter. 

Let $\mathbf{x}\in\mathbb{R}^{D\times T}$ denote the input signal consisting of the recordings /measurements from all the $D$ channels for $T$ time samples. 
Let us consider a GNN with parameters $\theta$ and let $z_\theta(\mathbf{x},A)$ denote the output of the last layer or the logits learnt by the GNN, where $A\in\mathbb{R}^{D\times D}$ denotes the graph between the channels. Further, let us use subscripts $t$ and $s$ for the teacher and student GNNs, respectively:  
 $z_{\theta_t}(\cdot,A)$ and $z_{\theta_s}(\cdot,A)$ denote the output layers from the teacher and student GNNs, respectively.
The teacher network is learnt by minimizing the following the binary cross entropy function $BCE(\cdot,\cdot)$ between the class label $y$ and the model prediction $f^t_{\theta_t}(\mathbf{x})$
\begin{equation}
    \mathcal{L}_{CE}(\theta_t)=\mathbb{E}_{\mathbf{x}}\left(BCE(y,z_{\theta_t}(\mathbf{x},A))\right),
    \label{eq:crossentropy}
\end{equation}
with respect to $\theta_t$, where $\mathbb{E}$ denotes the expected value obtained by averaging over all training samples $\mathbf{x}$. We use the BCE function since we consider here only the seizure versus non-seizure classification problem.
In order to train the student GNN from the pre-trained teacher, we minimize a regularized BCE cost, where the regularization term is given by the distillation loss that minimizes the KL divergence between the soft-output of the teacher and student GNNs:
\begin{align}
    \mathcal{L}_D(\theta_t*,\theta_s)=\mathbb{E}_{\mathbf{x}}\left(KL(p(z_{\theta_t*},T)(\mathbf{x},A),p(z_{\theta_s}(\mathbf{x},A),T))\right),
    \label{eq:distillation_loss}
\end{align}
where $\theta_t*$ denotes the parameters of the pre-trained teacher.
Then, the student network is trained by minimizing the total loss function:
\begin{equation}
    L_S(\theta_s)\triangleq(1-\delta)\mathcal{L}_D(\theta_t*,\theta_s)+\delta\,\mathcal{L}_{BCE}(\theta_s).
    \label{eq:totalloss}
\end{equation}
Our formulation so far uses the same input for both the student and teacher, and hence, the same number of input channels. This is because the KD formulation assumes that the input to both the student and the teacher are of the same class, as we discussed in the Preliminaries. However, our ultimate goal is to transfer knowledge to a student that uses the measurements from reduced set of nodes/channels $\mathbf{x}^d$ with $d<D$, and not $\mathbf{x}$. In other words, we wish to train a student model that works on a subgraph $A'$ of the original graph $A$. We achieve this by modifying the graph used by the student deleting the edges from the full graph with adjacency matrix $A$ as follows:
\begin{equation}
{A}' = W^\top\,A\,W,
\label{updated_adj}
\end{equation}
where $W\in\mathbb{R}^{D\times d}$ denotes the selection matrix which is a permutation of the matrix given by concatenation of a identity matrix of dimension $d$ with an all zero matrix of size $(D-d)\times d$ $-$ retains only the subgraph of $d$-size subset of the channels.\footnote{ In general, $A'$ may not necessarily be a connected graph, unless specifically regularized to be so.} The input $\mathbf{x}_d$ is then given by $\mathbf{x}_d=W^\top\mathbf{x}\in\mathbb{R}^{d}$, corresponding to the nodes of the subgraph defined by $W$. This in turn means that we must use $z_{\theta_s}(\mathbf{x}_d,A')$ and not $z_{\theta_s}(\mathbf{x},A)$ in the total loss function in \eqref{eq:totalloss}. Further, in order that the hidden nodes corresponding to the deleted channels are not pooled in the GNN, we multiple the output of each hidden layer of the GNN also by $W$. This in turn means that in practice the student GNN working on $D$ nodes can be fed with zeroes at the test time on the discarded channels, corresponding to having only the reduced set of measurement channels as input for seizure detection. We note that, while the specific application setting used in this work is that of scalp EEG channels, our proposed approach can be applied also to other multi-channel settings such as fMRI, where there is knowledge of connectivity across channels/measurements. The use of GNNs also makes our approach inherently interpretable in terms of connectivity of the brain regions. 

We consider three different instances of our model in this work:
(a)
    \textbf{G}lobal \textbf{S}tudent GNN with \textbf{P}re-\textbf{S}elected channel reduction (GS-PS) model, (b) \textbf{G}lobal \textbf{S}tudent GNN with \textbf{d}ata-\textbf{d}riven channel reduction (GS-DD) model, and
    (c) \textbf{P}ersonalized \textbf{S}tudent with \textbf{D}ata-\textbf{D}riven channel reduction (PS-DD) model
We describe them next.
\subsection{GS-PS Model}
\label{sec:gsps}
 We first consider the case when the reduced electrodes are preselected, or known already. In particular, we chose the four channels of T3, T4, T5, and T6 of the 19 channels from the T-20 montage \cite{jasper1958ten} as the reduced electrode set. This is motivated by input from neuroscientists that say these temporal channels can be relatively more indicative channels for seizure in general \cite{DBLP:journals/artmed/GabeffTZCRRA21}. In this case, the $W$ matrix from Eq. \eqref{updated_adj} corresponds to a diagonal matrix with ones only at the indices corresponding to T3, T4, T5, and T6. We also validate the choice of these channels through the following experiment.
We conduct an experiment where a new model with the same architecture as the teacher (keeping the full electrode channels) is trained to learn relevance weights $w$ for each electrode: this was simply achieved by applying a learnable diagonal matrix $M\in\mathbb{R}^{D \times D}$ to the input before the GNN such that the effective input to the GNN was defined as $\mathbf{x}_M' = M \cdot \mathbf{x}\in\mathbb{R}^{D\times D}$. We notice that the weights assigned to the temporal and some of the occipital electrodes were the highest, in particular, T2,T3,T4, and T5, were given large weights.
A more practical reason for the choice of temporal channels is the development of wearable sensors: many state of the art wearable sensors are of the behind the ear type, corresponding to these four temporal channels \cite{DBLP:journals/artmed/GabeffTZCRRA21,eglass}.
We apply the proposed GS-PS model for seizure detection by training them on the data from training patients and apply them to detect seizures on new test patients. In this case, the subgraph is pre-determined.
\subsection{GS-DD Model}
\label{sec:gsdd}
\begin{figure}[t]
  \vspace{-2mm}
  \centering
  \includegraphics[scale=0.35]{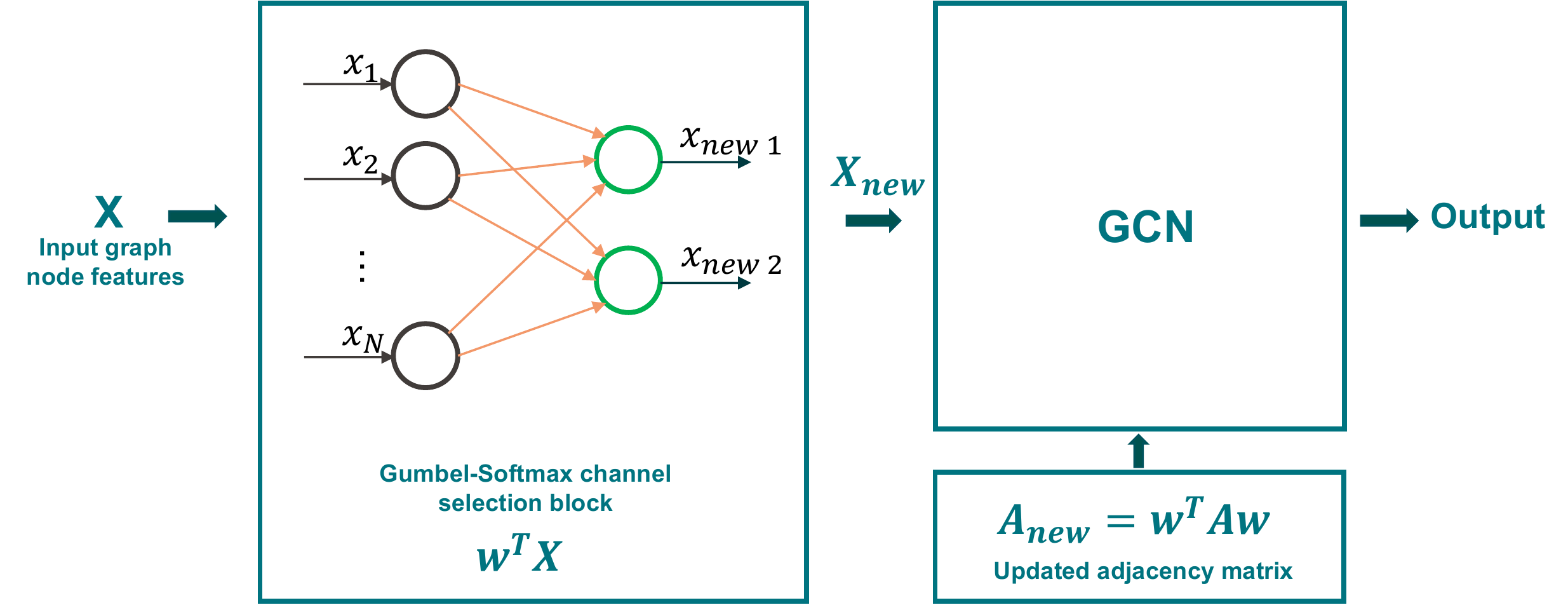}
  \caption{Proposed approach}
  \vspace{-6mm}
  \label{fig:GBSM_framework}
\end{figure}

We next consider the case of learning a student with channel reduction achieved in a completely data-driven manner. 
We propose to use a Gumbel-softmax (GS) channel selection block akin to the approach pursued in \cite{strypsteen2021end}. Our proposed GD-DD model consists of two connected blocks, first, the GS block that selects the subset of channels/electrodes, followed by the GNN block that produces a label as shown in Figure \ref{fig:GBSM_framework}. The details of the GS block are given next.

 The Gumbel-Softmax EEG channel selection block was first proposed by Strypsteen and Bertrand \cite{strypsteen2021end}, where channel selection was acheived through a learnable layer in the Deep Neural Network (DNN) in an end-to-end differentiable manner. The Gumbel-softmax layer represents a relaxation of the discrete selection operation that allows for differentiation \cite{strypsteen2021end} \cite{jang2016categorical} \cite{maddison2016concrete}. Let $x_n$ indicate the feature vector derived from channel $n$, and $x_{new_i}$ indicate the $i$th channel in the reduced set of channels. During training, the output of each selection neuron $k$ is given by $x_{new_k} = w_k^{T}X$, with $w_k$ sampled from the concrete distribution given by \cite{maddison2016concrete}:
\begin{equation}
\label{equa:GBSM_W}
    \centering
    w_{nk}=\frac{\exp((\log\alpha_{nk}+G_{nk})/\beta)}{\sum_{j=1}^{N}\exp((\log\alpha_{jk}+G_{jk})},
\end{equation}
with $G_{nk}$ independent and identically distributed samples from the Gumbel distribution and $\beta \in (0, +\infty)$ the temperature parameter of the concrete distribution. The effective subset of input node features is computed as $X_{new}=w^{T}X$.  The temperature parameter $\beta$ controls the extent of this relaxation from the one-hot selection: as $\beta$ approaches 0, the distribution becomes more discrete, the sampled weights converges to one-hot vectors. The continuous relaxation allows $w$ to be jointly optimized with model parameters, and to match the channel selection to the target model. The most pertinent EEG channels are thereby selected without prior expert knowledge or the need for manual feature selection. The learnable parameters $\alpha$ of this distribution are jointly optimized with  with the other network weights. At the end of training, the selection layer is made to select discrete channels by hard-thresholding the entries of $w_k$ so that they select only $K$ channels as $\displaystyle w_{nk}=
    \begin{cases}
      1 & \text{if $n={\arg\max}_{j}\alpha^{*}_{jk}$}\\
      0 & \text{otherwise},\\
    \end{cases}   $, 
where $\alpha^{*}$ is the learned matrix after training. We note that during test time, the GS block takes the form of a fixed linear matrix multiplication $W$ that acts to select the electrode channels. We also note that unlike the pre-selected case presented in Section \ref{sec:gsps}, GS-DD model learns a {\em data-driven subgraph}.
%
%

 In order to obtain a data-driven channel selection, we use the Gumbel softmax channel selection block as part of our GNN pipeline shown in Figure \ref{fig:GBSM_framework}. 
In particular, we apply the GNN on the reduced subgraph obtained by selecting only a subset of input EEG channel signals $X_{new}$ and that uses the adjacency matrix $A_{new}$ corresponding to the selected channels. 
As discussed above, the GS block is parameterized by a learnable matrix $\alpha \in \mathbb{R}^{N \times K}$, where $N$ is the total number of electrodes, and $K$ is the number of electrodes we wish to keep after reduction. When being fed a sample $X$, the selection block sample a weight matrix $W \in \mathbb{R}^{N \times K}$ from the concrete distribution following Equation \eqref{equa:GBSM_W}. This can be viewed as a softmax operation, which produces a weight matrix whose elements summing to one as continuous relaxation of one-hot vectors. In our experiments, we use a similar method as in paper \cite{strypsteen2021end}. During the training, we set $\beta(t)=\beta_{s}(\beta_{e}/\beta_{s})^B$, decreasing in an exponential manner where $B$ is the total number of training epochs. In particular, $\beta(t)$ is the temperature parameter at epoch $t$, $\beta_{s}$ and $\beta_{s}$ are respectively the starting and ending $\beta$. In our settings, $\beta_{s}=100$, $\beta_{s}=0.001$. As we noted before, while the complete set of electrodes is indeed used during training of the student GNN, this is not the case during test time as the $W$ matrix will be set to ones and zeros, thereby not requiring any measurements from the non-selected electrodes.

\textbf{Channel consolidation}
We note that, though we force the weight matrix to select a reduced set of channels, it is possible that a given channel is chosen multiple times since we have not actively enforced that there is no duplication. In order to discourage duplicate channels, we minimize the total loss regularized with the penalty given by \cite{strypsteen2021end}: $\displaystyle\Omega(P)=\lambda \sum_{n=1}^{N}\mathrm{ReLU}(\sum_{k=1}^{K}p_{nk}-\tau)$, 
where $\mathrm{ReLU}(\cdot)$ is the rectified linear unit, $\lambda$ is the weight of the regularization loss, and $\tau$ the threshold parameter. During training, we set $\tau(t)=\tau_{s}(\tau_{e}/\tau_{s})^B$, decreasing in an exponential manner. In our settings, $\tau_{s}=3$, $\tau_{s}=1.1$. $\lambda$ is set to be $5$ to control the strength of the regularization.

Then, we learn the GS-DD model with the regularized student loss, to obtain a seizure detection model that is global and applicable to any patient.

\subsection{PS-DD model}
\label{sec:personalized_seizure_detection}
Epileptic seizures vary significantly between individuals and personalized models could be beneficial in taking into account their unique patterns and characteristics. This motivates us to extend our previous model to a personalized setting to for simultaneous electrode reduction and seizure detection for every single patient. 
As with the GS-PS and GS-DD models proposed in Sections \ref{sec:gsps} and \ref{sec:gsdd}, our aim here is to arrive at light-weight models for seizure detection that use only a subset of electrode channels using KD, but personalized to the patient. As with GS-DD model, we let the channels be selected in a data driven manner. 
 Our hypothesis is that {\em both knowledge-distillation and personalized models have an important role} to play in improving the seizure detection performance, {\em particularly in the cases when the available data is scarce.}
The PS-DD model is in its essence the same as the GS-DD model in the architecture, with the crucial difference that the model is now trained in a patient-specific manner. This means that the PS-DD model also learns a {\em data-driven subgraph for every patient}.

\section{Numerical Experiments}
\label{sec: numerical_experimental}
\subsection{Settings}
\textbf{Dataset} We apply our models for the task of seizure detection on the data from the Temple University Hospital EEG Seizure Data Corpus (TUSZ) \cite{dataset}, which is one of the most popularly used, and diverse datasets that consists of over 100 patients of a wide range of ages ($8$-$80$) with different seizure types, e.g., focal seizure, tonic seizure, generalized non-specific seizure, and myoclonic seizure for long durations. The data is in the form of 19 channel EEG recordings in the 10-20 electrode placement system. As our work deals with the problem of seizure detection, no distinction is made between seizure types and all seizures were grouped into one class, resulting in a binary classification problem. The selected seizure (ictal) segments ranged between 5 and 60 seconds in length. Corresponding interictal segments of the same length were selected that ended one minute before seizure onset, following the methodolgy pursued in \cite{DBLP:journals/artmed/GabeffTZCRRA21}. This resulted in a balanced dataset of 50\% seizures and 50\% nonseizure segments. The segments are taken sequentially without overlap. Every segment is then divided into windows of 5 seconds for both the classes. All selected segments were then split into five-second windows. The TUH dataset has two separate sets of recordings for \texttt{train} and for \texttt{dev}, which correspond to different set of patients for training and test, respectively. Similarly to the literature, we use only the patients from \texttt{train} for training models, and the test patients from \texttt{dev} for testing the learnt models on which the performance is reported. Finally, we have a total of  14382 samples for training and 4529 samples for testing, each sample being a 5 second window of multi-channel EEG signal.

\textbf{Data preprocessing}
As customary in EEG signal processing, each sample is then filtered with a Butterworth bandpass filter of order 5 between 0.5 and 50 Hz to remove the artifacts and noise. Similarly to \cite{ensemble-eeg}, the features were calculated for each EEG channel: 
energy of the signal filtered in frequency bands (from 0.5 Hz to 30 Hz with a bandwidth of 3 Hz and from 30 Hz to 50 Hz with a bandwidth of 10 Hz), Hjorth complexity and mobility, decorrelation time, L2-norm of the approximation and detail coefficients obtained from a six-level wavelet decomposition using a Daubechies db4 wavelet, log amplitudes of the non-negative frequency components. 
This results in 647 features in total for each sample/window. The features are then normalized component-wise and taken as input $\mathbf{x}$ to the GNN along with the distance based adjacency matrix.

\textbf{Training data}
In order to train the teacher, no distinction is made between patients or segments and the entire training data is used to train the teacher.
All the samples from all the test patients are used as test data.
For the training the global models of 
GS-PS, and GS-DD, we use the data of all training patients during training and data from all test patients for testing. On the other hand, since the PS-DD model is trained for each patient separately, the training and test data segments are obtained by splitting the {\em segments} of the given patient randomly. Further, in order to understand the effect of personalization, we divide the patients into three bands based on the amount of data segments they possess as shown in Table \ref{tbl:bands}.

\begin{table}[t]
\vspace{-.1in}
\centering
\caption{\label{tbl:bands}Three bands of patients.}
\begin{tabular}{ccccl}
\toprule
\textbf{Data Bands} &\quad \textbf{Number of Segments N} &\quad \textbf{Batch Size} & \quad\textbf{Epoch} &  \\ 
\midrule
Rare Data     & $4 \leq N <20$                & 2                   & 20             &  \\
Mid Data      & $20 \leq N <100$              & 16                  & 100            &  \\
Rich Data     & $ N \geq 100$                 & 64                  & 100            &  \\ 
\bottomrule
\end{tabular}
\vspace{-.1in}
\end{table}

\textbf{Model Training}
\label{subsec:modeltraining}
We use a two-layer GCN network with $32$ hidden nodes in each hidden layer as the teacher model. It is trained with a batch size of 64 and a learning rate of $10^{-5}$. The student network in all the three cases of GS-PS, GS-DD, and PS-DD, is a light-weight model with just one-layer GCN of only 1 hidden node. We note that the number of parameters to learn in the student is {\em just $3\%$} of that of the teacher. Each of the three models is trained and tested both with and without KD in order to determine the contribution of the teacher knowledge. As described in Equation \eqref{eq:totalloss}, the KL divergence loss is used as the distillation loss function and the binary cross-entropy loss is used for the student loss function. The hyperparameters in the total loss are obtained by performing $5$-fold cross-validation. We set $T=5$, and $\delta$ values are set to $0.1, 0.5, 0.8$. For GS-DD, we consider the case of $K=4$ channels to compare the performance with that of GS-PS using the four temporal channels. For the PS-DD model, we use $K=2$ electrodes for every patient.
 
\textbf{Evaluation metrics}
\label{sec:expNrslt}
Following \cite{covert2019temporal,seizurenet}, we evaluate the performance of the three models using two standard metrics: f1-score and the Area Under the curve of the Receiver Operating Characteristic (AUROC), which are used regularly in evaluating seizure detection. In all the cases, the performance is averaged over the different test patients.
\vspace{-.1in}
\subsection{Detection Performance results}
We now report the performance of the different approaches.

\textbf{GS-PS model}
The performance of the teacher and the global student with the pre-selected temporal channels is presented in Table \ref{tbl:GCN_result_ingeneral}. In the pre-selected student, we observe that KD significantly improves the performance in terms of the f1-score that tends to be comparable to that of the teacher.

\textbf{GS-DD model}
 Unlike in the temporal channel pre-selection case, we see that the performance remains relatively constant to the different levels of KD. This is probably because the GS selection already results in a high performance, and the teacher does not offer notable improvement. 

\begin{figure}[t]
\begin{minipage}[b]{0.5\linewidth}
\centering
  \includegraphics[scale=0.375]{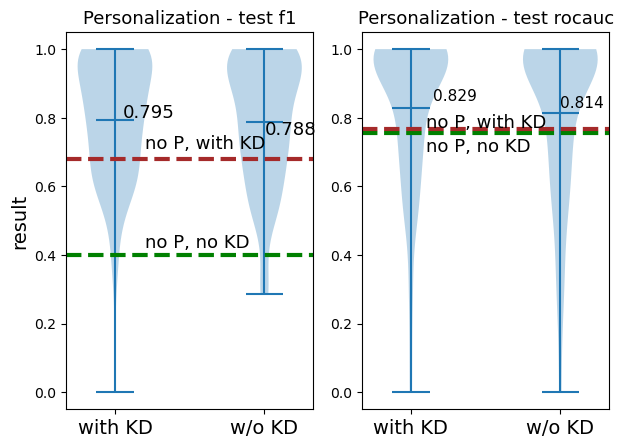}
  \caption{The effectiveness of personalization: With personalization (P), both test f1 and test rocauc are significantly improved on average.\\}
\label{fig:personalization}
\end{minipage}
\hspace{0.08cm}
\begin{minipage}[b]{0.45\linewidth}
\centering
  \includegraphics[scale=0.38]{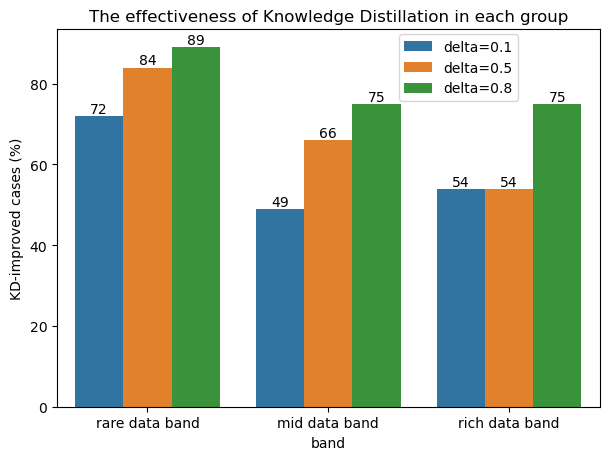}
  \caption{The effectiveness of Knowledge Distillation (KD) in different bands\\}
\label{fig:barKD}
\end{minipage}
\vspace{-.25in}
\end{figure}

 \textbf{PS-DD model}
In the case of a personalized student GNN with only two electrodes (that we call PS-DD 2), we observe that the performance improves as $\delta$ is increased, meaning more emphasis is given to the patient's data over the teacher's knowledge, with the highest performance obtained at $\delta=0.8$. On the other hand, we also observe that completely relying on the patient's data and not using the teacher ($\delta=1$) reduces the performance. Further, we note that the performance of the student even without teacher's knowledge ($\delta=1$) is generally much better than that of the teacher or the global student. This in turn supports our intuition and hypothesis that personalization also plays a significant role in improving seizure detection performance.
In the two plots in Figure \ref{fig:personalization}, we depict the distributions of test f1 and AUROC of all test patients in the circumstances with or without KD, respectively for the PS-DD model. The averaged performances are indicated in numbers in the figures. The dashed red/green lines show the general performances of models without personalization.  When trained on the general population, we obtain the test f1 of models with and without KD as $0.7$ and $0.4$, respectively. Whereas after personalization, the average test f1 are improved by $16\%$ and $50\%$  to around $0.8$, corresponding to with and without KD, respectively. This shows that by tackling the diversity in EEG seizure data on a large population, personalization has the potential to improve seizure detection. The average test AUROC is improved by $8\%$ to above $0.8$. The detailed results are reported in Table \ref{tbl:GCN_result_ingeneral}. However, the average performance with KD is only slightly higher than the average performance without KD in both the metrics. This in turn motivated us to look into the performance in the three data bands individually next.
\begin{table}[t]
\vspace{-.0in}
  \caption{Test Results with Different Models}
  \setlength{\tabcolsep}{0.7\tabcolsep}
  \centering
  \begin{adjustbox}{width=1\textwidth}
  \begin{tabular}{ *{10}{c} }
    \toprule
    \multicolumn{2}{c}{\textbf{Model}} & \multicolumn{2}{c}{\textbf{w/o KD}} & \multicolumn{6}{c}{\textbf{w/ KD}}\\
    \cline{1-10} \\
    \text{Channel} & \multirow{2}{*}{\text{Personalization}} & \multicolumn{2}{c}{--} & \multicolumn{2}{c}{$\delta=0.1$} & \multicolumn{2}{c}{$\delta=0.5$} & \multicolumn{2}{c}{$\delta=0.8$} \\
    \cline{3-10} \\
    \text{Selection} &                                       &  f1 & auroc &  f1 & auroc  &  f1 & auroc  &  f1 & auroc  \\
    \toprule
    \text{Teacher} & {--} & 0.689  & 0.781 & {--} & {--} & {--} & {--} & {--} & {--}\\
    
    \multirow{1}{*}{\text{GS-PS}} & \multirow{1}{*}{$\times$} & 0.401  & 0.755 & 0.683 & 0.766 & {--} & {--} & {--} & {--} \\
    \multirow{1}{*}{\text{GS-DD}} & \multirow{1}{*}{$\times$} & { }0.690{ }  & { }0.763{ } & 0.695  & 0.761 & 0.697  & 0.763 & 0.693  & 0.764 \\
    \multirow{1}{*}{\text{PS-DD}} & \multirow{1}{*}{$\checkmark$} & 0.788  & 0.814 & { }0.755{ }  & { }0.777{ } & { }0.784{ }  & { }0.829{ } & { }\textbf{0.795}{ }  & { }\textbf{0.829}{ } \\

    \bottomrule
  \end{tabular}
  \end{adjustbox}
  \label{tbl:GCN_result_ingeneral}
\end{table}

\subsection{Performance analysis}
\vspace{-.1in}
\label{subsec: KDeffectBands}
To better understand the effectiveness of our models, we do a detailed  performance analysis by further dividing patients into three bands based on the number of seizure segments (rare-data band, mid-data band, rich-data band) and delve into the performances, respectively as shown in Table \ref{tbl:bands}.
In Table \ref{tbl:band_compare_all}, we report the seizure detection results when the model training relies differently on the new patient data to different levels given by $\delta=0.1$, $\delta=0.5$ and $\delta=.8$, respectively, in \eqref{eq:totalloss}. The setting of $\delta = 0.8$ corresponds to the case where the student training relies more heavily on unseen patient-specific data than the teacher. Figure \ref{fig:barKD} shows the differences in the percentages of cases in each band where KD boosted the model performance (in terms of test f1 and test AUROC). Overall, KD helps $72\%$ (47 out of 65) patients in the rare-data band improve their model testing performances. But only $49\%$ (26 out of 53) patients in the mid-data band and $54\%$ (18 out of 33) patients benefit from the teacher. In general, we observe the tendency that patients with scarce data benefit the most from KD. This gives us the motivation to further delve into the rare-data band case. 
%

In the rare-data band, we notice that we constantly encounter four patients with the lowest performance that bias the overall performance significantly. It turns out that these four cases correspond to the patients with the least training data. We refer to these cases as the four "extremes" in our experiments. Since the TUSZ dataset is rather diverse and we wish to see the averaged performance without a strong bias, we chose to exclude the extremes out and recompute the performance metrics. We notice from Table \ref{tbl:exclude_extreme}, that the performance improves overall by excluding the extremes, and the best performance is obtained when $\delta=0.8$. This indicates that the effectiveness of KD in personalized settings widely varies with the amount of data each patient possesses, and potentially across the patient types (since the dataset includes different types of seizures that we do not currently account for) and also varies with the change of the weight of student loss $\delta$. In our experiments, $\delta$ is 0.8 gave the best scores on an average. A more exhaustive approach would be to compute personalized models with personalized $\delta$, but that is beyond the scope of the current work.
%
\begin{table}[t]
  \caption{PS-DD Test Results on Different Bands}
  \setlength{\tabcolsep}{0.7\tabcolsep}
  \centering
  \begin{adjustbox}{width=1\textwidth}
  \begin{tabular}{ *{10}{c} }
    \toprule
     & & \multicolumn{2}{c}{\textbf{w/o KD}} & \multicolumn{6}{c}{\textbf{w/ KD}} \\
    \cline{3-10} 
    \multirow{2}{*}{\textbf{Data Bands}} & \multirow{2}{*}{\textbf{{ }Personalization{ }}} & \multicolumn{2}{c}{\textbf{--}} & \multicolumn{2}{c}{\text{$\delta=0.1$}} & \multicolumn{2}{c}{\text{$\delta=0.5$}} & \multicolumn{2}{c}{\text{$\delta=0.8$}} \\
    \cline{3-10} \\
             &         & f1 & auroc & f1 & auroc & f1 & auroc & f1 & auroc \\
    \midrule
    \text{Rare Data} & \multirow{1}{*}{$\checkmark$}& 0.786 & { }0.791 & { }0.783 & { }0.783 & { }$0.790^*$ & { }$0.816^*$ & { }\textbf{0.798*} & { }\textbf{0.827*} \\
    \text{Mid Data} & \multirow{1}{*}{$\checkmark$}& \textbf{0.791} & { }\textbf{0.837} & { }0.726 & { }0.756 & 0.774 & 0.819 & 0.786 & 0.833 \\
    \text{Rich Data}&\multirow{1}{*}{$\checkmark$} & 0.790 & { }0.821 & { }0.749 & { }0.800 & 0.786 & \textbf{0.829} & { }\textbf{0.801*} & { }$0.828^*$ \\
    
    \bottomrule
  \end{tabular}
  \end{adjustbox}
  \label{tbl:band_compare_all}
\end{table}
\begin{figure*}[t]
  \vspace{-.1in}
  \centering
  \includegraphics[width=\textwidth,height=4.3cm]{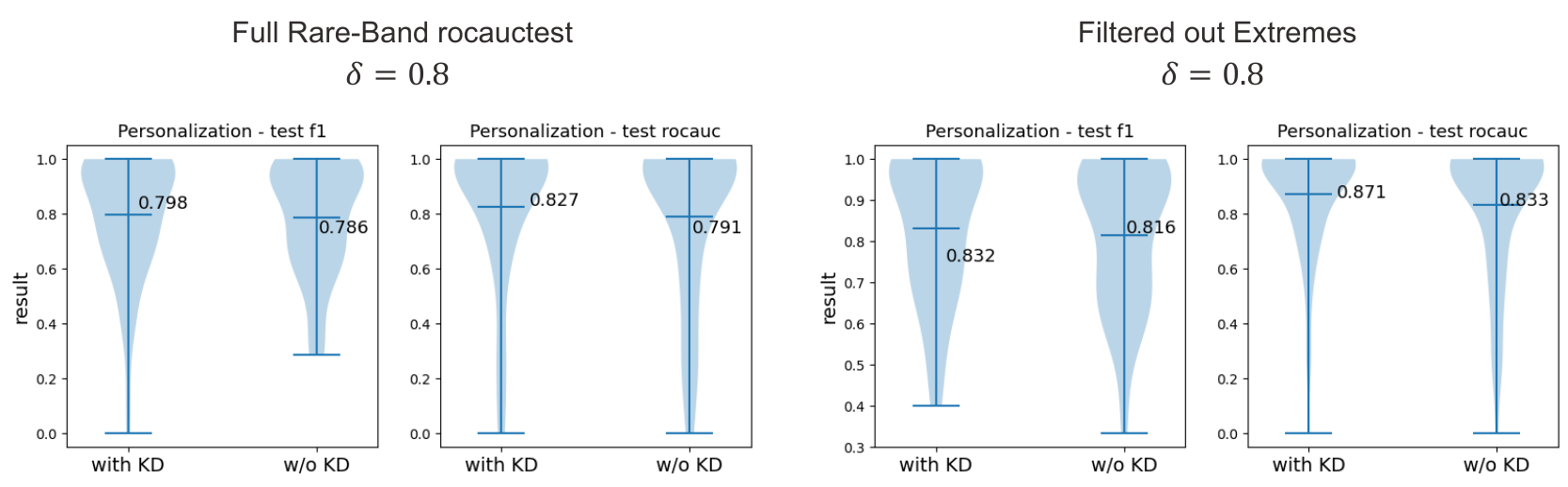}
  \caption{PS-DD Test Results ($\delta = 0.8$) on Rare-Data Band. The right column shows the results with 4 patients with extremely sparse data and poor performances excluded (Patient ID: ”00005672”, ”00008706”, ”00006535”, 00004596”)\\}
  \vspace{-6mm}
  \label{fig:rare_compare}
\end{figure*}
%
%
 \begin{table}[t]
  \caption{\label{tbl:exclude_extreme}PS-DD Test Results on Rare-Data Bands}
  \setlength{\tabcolsep}{0.7\tabcolsep}
  \centering
  \begin{tabular}{ *{6}{c} }
    \toprule
    \multicolumn{2}{c}{\multirow{1}{*}{\textbf{Model}}} & \multicolumn{2}{c}{\multirow{1}{*}{\textbf{Extremes $\times$}}} & \multicolumn{2}{c}{\multirow{1}{*}{\textbf{Extremes 
    $\checkmark$}}}\\
    \midrule
    $\delta$ & \text{{ }Personalization{ }} & f1 & auroc & f1 & auroc \\
    \midrule
    0.1 & \multirow{1}{*}{\text{$\checkmark$}} & { }0.794{ } & { }0.806{ } & { }0.783{ } & { }0.783{ }\\
    0.5 & \multirow{1}{*}{\text{$\checkmark$}} & { }0.813{ } & { }0.860{ } & { }0.790{ } & { }0.816{ }\\
    0.8 & \multirow{1}{*}{\text{$\checkmark$}} & { }\textbf{0.832}{ } & { }\textbf{0.871}{ } & { }0.798{ } & { }0.827{ }\\
    1 (no KD) & \multirow{1}{*}{\text{$\checkmark$}} & { }0.816{ } & { }0.833{ } & { }0.786{ } & { }0.791{ }\\
    
    \bottomrule
  \end{tabular}
  \vspace{-.1in}
\end{table}

\textbf{Effectiveness of Knowledge Distillation when lacking informative channels/signals}
\label{subsec: KDeffectChannels}
%
To further test the effectiveness of both personalization and KD, we select to keep we arbitrarily select to keep only signals from channels FP1, FP2 that belong to the frontal region, which are suggested to be the less informative region for epileptic seizure detection.
The Gumbel-Softmax channel selection block is not involved in this section. The experiment is conducted on the rare data band, with the hypothesis that the combination of personalization and KD can help compensate for the adverse situation brought by a) lack of data, and b) lack of informative channels.
With only personalization but no KD, 53.8\% (35 out of 65) patients' test f1 and AUROC score still exceed 0.65, yielding fairly good performances. In the rest of not the ideal personalized situations, 90\% (27 out of 30 patients) benefit from the teacher. With even the alleged least informative channels, we get 53.8\% of the cases with rather promising results. For the rest of the cases, the integrated application of personalization and KD has been observed to be effective for detecting epileptic seizures. We thus see that the combination leverages the strengths of both techniques to provide highly accurate results in scarce data scenarios.

\begin{figure}[t]
  \vspace{-2mm}
  \centering
  \includegraphics[scale=0.4]{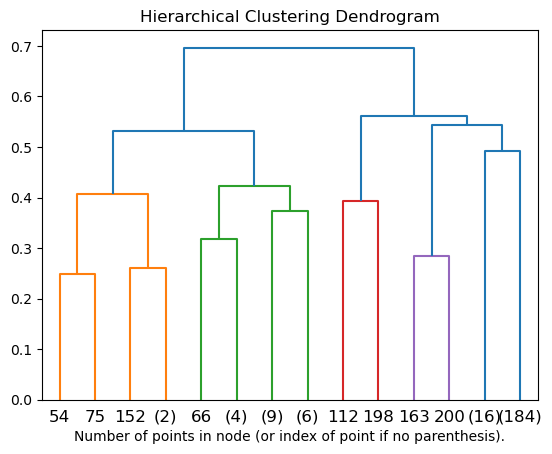}
  \caption{Hierarchical clustering of patients based on their Gumbel-Softmax channel selection patterns\\}
  \vspace{-6mm}
  \label{fig:cluster}
\end{figure}

 %
 %
\textbf{Hierarchical Clustering of Patients}
\label{subsec:clustering}
We now investigate if the different patients naturally show clusters when the learnt electrode channels are used to cluster the patients. We use hierarchical clustering on the learnt selection matrices $W$. Hierarchical clustering is a method of cluster analysis that builds a hierarchy of clusters by successively splitting or merging smaller clusters based on the Euclidean distance between clusters.
The result of hierarchical clustering is shown as a dendrogram that shows the hierarchy of clusters in Figure \ref{fig:cluster}. We observe that there are no clearly significant clusters emerging except for a large cluster and outliers, which could be because the patients and seizure signals in the TUSZ dataset are quite diverse. We also note that we have made no distinction between seizure types (about 6 of widely varying number of samples per type) in our analysis which might explain the single big cluster. While some of the outliers corresponded to patients with rare disease (Rasmussens' syndrome), it is unclear if the outliers show specific signature characteristics that separate them clinically from the main cluster. Further, we see that the main cluster diameter is relatively large indicating that there is significant variability in the selected channels across the different patients. In future work, we plan to pursue alternative clustering strategies and features and also mitigating the diversity by, for example, filtering out only the focal seizure signals.

\vspace{-.1in}
\section{Conclusions and future work}
We proposed an approach to transfer the knowledge from a pre-trained GNN-based seizure detection to the case when the number of measurement electrodes is reduced. We showed that it is possible to obtain models that are  (i) light-weight (requiring just a $3\%$ of the sophisticated network), and (ii) work on reduced electrodes (requiring as low as only $10\%$ of the original electrodes), yet offer superior performance in seizure-detection, particularly in the personalized setting. The approach resulted in patient-specific choice of the reduced set of electrodes. Our experiments demonstrated the merit of both knowledge distillation and personalization, particularly when dealing with patients with scarce data. We observe that there is a trade-off between the use of prior information (teacher) and patient-specific data: although teacher-knowledge is necessary, the relative importance should be higher on the patient-specific data for maximum performance. We believe that these results show that our approach can provide meaningful insights and guidelines in the practical setting where there is need to move from full scalp electrode measurements to reduced form factor measurements, such as personalized wearable devices. 
We have currently restricted our analysis to a relatively simple GNN teacher model and used the graph given by physical placement of electrodes. The quality of the teacher and the graph used both translate into the quality of the student model, and hence, we believe that a more sophisticated GNN could be employed to further improve overall performance. In the future, it would also be interesting to look into multi-class seizure classification and identify the different types of seizures.

\clearpage
\bibliographystyle{splncs04}
\bibliography{literature}

%
%
%
%

\end{document}